\documentclass[printer]{tMOP2e}
\usepackage{epsfig}
\usepackage{bm}
\usepackage{amsmath}

\begin{document}

\title{Speed-up and slow-down collisions in laser-induced nonsequential multiple
ionization}
\author{C. Figueira de Morisson Faria$^*$\thanks{$^*$Corresponding author. E-mail: c.f.m.faria@city.ac.uk}$^{1}$ and X. Liu$^{2}$ \\
$^{1}$Centre for Mathematical Science, City University, Northampton Square,
London EC1V OHB, UK\\
$^{2}$Department of Physics, Texas A\&M University, College Station, TX
77843-4242, USA}
\date{\today}
\maketitle

\begin{abstract}
We investigate laser induced nonsequential multiple ionization using
a simple statistical model in which an electron recollides
inelastically with its parent ion. In this collision, it thermalizes
with the remaining $N-1$ bound electrons within a time interval
$\Delta t$. Subsequently, all the $N$ electrons leave. We address
the question of how the above time delays influence the individual
contributions from the orbits in which the first electron, upon
return, is accelerated or decelerated by the field, respectively, to
the ion momentum distributions. In both cases, the time delays
modify the drift momenta obtained by the $N$ electrons when they
reach the continuum at $t+\Delta t$, by moving such times towards or
away from a crossing of the electric field. The contributions from
both types of collisions are influenced in distinct ways, and the
interplay between such trends determines the widths and the peak
momenta of the distributions. Specifically in the few-cycle pulse
case, we also show that such time delays do not affect the shapes of
the momentum distributions in a radical fashion. Hence, even with a
thermalization time $\Delta t$, nonsequential multiple ionization
could in principle be used for absolute-phase diagnosis.
\end{abstract}

\setcounter{MaxMatrixCols}{10}

\markboth{Speed-up and slow-down collisions in laser-induced
nonsequential multiple ionization}{C. F. de M. Faria and X. Liu}
\section{Introduction}

The physics of non-sequential double ionization in rare gases by
low-frequency, intense laser fields is, to a very large extent, understood.
In particular for neon, the electron momentum distributions with peaks at
non-vanishing momentum components $p_{1||}=p_{2||}=\pm 2\sqrt{U_{p}}$, along
the laser-field polarization, or the ion momentum distributions peaked at
the parallel momenta $P_{||}=\pm 4\sqrt{U_{p}},$ where $U_{p}$ denotes the
ponderomotive energy, can be explained by a laser-induced inelastic
recollision process \cite{Ffm2000,FrMBI2000,AAMOP}. Specifically, the first
electron leaves an atom by tunneling ionization at a time $t^{\prime }$,
propagates in the continuum, being accelerated by the field and,
subsequently, at a time $t,$ recollides with its parent ion. The second
electron is then freed by electron-impact ionization \cite{corkum}.

This process has been successfully dealt with using an S-Matrix approach and
its classical counterpart \cite{faisal,KBRS,FFetal04,FFetal04R,fewcycle2}.
In particular, if the laser-field intensity is high enough, both
quantum-mechanical and classical models yield very similar momentum
distributions\footnote{%
This, however, no longer holds if the intensity is near threshold, i.e.,
just enough for the second electron to be released. For a detailed
discussion c.f. \cite{FLB2005}, and for measurements of nonsequential double
ionization below the threshold see, e.g., \cite{Ne03}.}. We have shown that
this agreement occurs for several types of electron-electron interaction,
and in the absence or presence of Coulomb repulsion in the electron final
states \cite{KBRS,FFetal04,FFetal04R}. Furthermore, we have verified that
the outcome of the classical and quantum mechanical computations also
exhibit a striking agreement if the driving field is a few-cycle laser pulse
\cite{fewcycle1,fewcycle2}. In particular, both the classical and quantum
mechanical distributions are strongly dependent on the so-called absolute
phase, i.e., the phase difference between the pulse envelope and its carrier
oscillation. We have used this fact in order to propose a scheme for
diagnosing the absolute phase \cite{fewcycle1}, which has also been
experimentally realized \cite{fewcycleexp}.

If one is dealing, however, with more than two active electrons, there are
many open questions concerning the physics behind laser-induced multiple
ionization. This is in particular true for triple and quadruple ionization
of neon by near infra-red ($\omega =0.057$ a.u.) laser fields in an
intensity range below $2\times 10^{15}$ $\mathrm{W/cm}^{2}$ \cite
{FrMBI2000,multi1,multi2}. The distributions observed in such experiments,
approximately peaked at non-vanishing ion momenta $P_{||}=\pm 2N\sqrt{U_{p}}$%
, where $N$ denotes the number of electrons involved, hint at a
non-sequential physical mechanism. The theoretical modeling of
non-sequential multi-electron processes for $N>2$, however, poses a far
greater challenge than in the two-electron case. For instance, already for
three electrons, which is the simplest scenario to be taken into
consideration, one must include at least six Feynman diagrams when computing
the corresponding transition amplitude, each of which is rather cumbersome
to implement.

In order to tackle such a problem, in previous work \cite{prltherm}, we
constructed a simple thermalization model, similar to those employed in, for
instance, nuclear \cite{hagedorn} or molecular \cite{RRKM} physics. We
assumed that the first electron, upon return, recollides with its parent ion
and shares its kinetic energy with $N-1$ bound electrons. All the $N$
electrons are then freed after a time interval $\Delta t$ subsequent to the
recollision time $t$, which, to first approximation, is taken to be
constant. We regard this time as the upper bound for a thermalization time,
which is necessary for the kinetic energy to be redistributed among the $N$
electrons.

The above-mentioned process takes place within a fraction of the cycle of
the laser field. Since the duration of a typical titanium-sapphire pulse,
employed in such experiments, is of the order of 2.7 $fs$, $\Delta t$ falls
within the attosecond regime. Indeed, recently, by comparing our
thermalization model with the available experimental data for triple and
fourfold ionization of neon \cite{multi1,multi2}, we have been able to
estimate thermalization times of the order of 460 attoseconds \cite{prltherm}%
. Using such a time delay, we computed ion momentum distributions whose
widths and peak momenta agreed very well with those in the experiments.

To a very large extent, our thermalization model contains classical
ingredients and is a generalization of the classical model already employed
in the two-electron case \cite{FFetal04R,FFetal04}. In \cite{prltherm}, we
have also observed that the peaks and the width of the momentum
distributions depend very strongly on the time delay $\Delta t.$ The
position of the peaks is a direct consequence of the kinematics involved in
the process, as the final drift momentum the $N$ electrons acquire from the
field, will strongly depend on the time $t+\Delta t$ when they reach the
continuum.

The above-stated physical mechanism, however, is not completely agreed upon.
Indeed, it could well be \ that a broadening in the electron momentum
distributions and the displacement in the peak momenta are caused by, for
instance, partially sequential ionization processes \cite{multi2}, or
excitation tunneling mechanisms \cite{sepmech} , in which, during the
recollision, the electrons undergo a transition to an excited state, from
which they subsequently tunnel (for an example of such mechanisms for
different atomic species see, e.g., \cite{ArvsNe}). Specifically for large
thermalization times, the distributions are very much concentrated near
vanishing parallel momenta, and resemble those obtained with
excitation-tunneling. Apart from that, even assuming that thermalization
through electron-impact ionization is the only or at least the dominant
physical mechanism, it is not clear how the time delay $\Delta t$ affects
the width of the distributions.

Another open question concerns the influence of the types of collisions on
the momentum distributions. In fact, for the recombination or rescattering
phenomena occurring in the context of atoms in strong laser fields, there
exist two main orbits along which an electron may return to its parent ion
with the same kinetic energy. Such orbits are known as ``the long orbit'',
or ``the short orbit'', and have been extensively investigated in the
context of high-order harmonic generation and above-threshold ionization
\cite{lshort} and, more recently, for nonsequential double ionization \cite
{panfili2002,FB2003}. An electron following the long orbit leaves the atom
at a time $t^{\prime }<0.3T $, where $T=2\pi /\omega $ denotes a field
cycle, and returns after the minimum of the electric field, whereas an
electron along the short orbit reaches the continuum at $t^{\prime }>0.3T$
and returns before the minimum of the electric field. Hence, in the former
and in the latter case the returning electron is decelerated and accelerated
by the field, respectively.

In recent years, both orbits have been investigated in the context of
non-sequential double ionization using classical ensemble models \cite
{panfili2002}. In particular, they have been associated to two main types of
collisions. The orbits for which the electron is being decelerated have been
related to the so-called ``slow-down collisions'', whereas the short orbit
has been linked to the so-called ``speed-up collisions''. Depending on
several parameters, such as the field intensity or frequency, one type of
collisions may provide the dominant contributions to the yield, or both
types may be equally important \cite{china}. For instance, for the
parameters in \cite{panfili2002}, it was shown that the slow-down collisions
dominate.

In this paper, we address the question of how the contributions from
the short and the long orbit to the multiple-ionization momentum
distributions are affected by the time delay $\Delta t$, for triple
and fourfold ionization. We are mainly concerned with the influence
of both sets of orbits on the width and the peaks of the ion
momentum distributions, for resolved and non-resolved ion momentum
components transverse to the laser-field polarization. We take the
driving field to be a monochromatic wave or a few-cycle pulse, and
make a detailed analysis of the similarities and differences between
both cases. This manuscript is organized as follows: In the
subsequent section, we provide a brief discussion of our
thermalization model, with the explicit expressions for the ion
momentum distributions. In Sec.~\ref{results}, we display our
results for the different sets of orbits, together with the total
yields and, finally, in Sec.~\ref{conclusions}, we state our
conclusions.

\section{Model}

\label{model} We will briefly recall the statistical model employed in our
previous publication \cite{prltherm}, in order to determine the momentum
distributions in N-fold non-sequential ionization. This model is mainly
composed by three main ingredients.

Firstly, we assume that one electron tunnels into the continuum at an
instant $t^{\prime }$ according to the quasi-static rate \cite{LL}.
\begin{equation}
R(t^{\prime })\sim |E(t^{\prime })|^{-1}\exp \left[
-2(2|E_{01}|)^{3/2}/(3|E(t^{\prime })|)\right] ,  \label{quasistatic}
\end{equation}
where $E(t^{\prime })$ and $|E_{01}|$ denote the electric field at the time
the electron left and the first ionization potential, respectively.

Subsequently, this electron propagates under the sole influence of the laser
field, and obeys the classical equations of motion. At a later time $t$, it
may be driven back towards its parent ion, recolliding inelastically with
it. The return condition $\mathbf{r_{1}}(t)=0$ for the first electron yields
\begin{equation}
F(t)=F(t^{\prime })+(t-t^{\prime })A(t^{\prime }),  \label{tangent}
\end{equation}
where $\mathbf{A}(t)$ is the vector potential, and $F(t)=\int^{t}{A(\tau
)d\tau }$. Eq. (\ref{tangent}) provides a clear geometrical picture for the
return condition: for a given $t^{\prime }$, the return time $t$ is given by
the intersection of $F(t)$ with its tangent at $t^{\prime }$ \cite{tgpapers}%
. We will employ this construction in Sec. \ref{results}, when discussing
our results (c.f. Fig.~2).

Upon recollision, the energy $E_{\mathrm{ret}}(t)=[A(t)-A(t^{\prime
})]^{2}/2 $ of the returning electron is completely thermalized among such
an electron \emph{and} the $N-1$ electrons to be freed. After the time
interval $\Delta t $, the distribution of energy and momentum over the $N$
electrons is only governed by the available phase space. We will consider
here that the time delay $\Delta t$ is constant. Finally, at the time $%
t+\Delta t$, the $N$ electrons reach the continuum with the total kinetic
energy $E_{\mathrm{ret}}-E_{0}^{(N)}$. The quantity $E_{0}^{(N)}=%
\sum_{n=2}^{N}|E_{0n}|$ denotes the total ionization potential of the $N-1$
(up to the recollision time $t$ inactive) electrons.

The distribution of final electron momenta $\mathbf{p}_{n}\ (n=1,\dots ,N)$
is proportional to
\begin{eqnarray}
F(\mathbf{p}_{1},\mathbf{p}_{2},\dots ,\mathbf{p}_{N}) &=&\int dt^{\prime
}R(t^{\prime })\delta \left( E_{0}^{(N)}-E_{\mathrm{ret}}(t)\right.  \notag
\\
&&\left. +\frac{1}{2}\sum_{n=1}^{N}[\mathbf{p}_{n}+\mathbf{A}(t+\Delta
t)]^{2}\right) ,  \label{F}
\end{eqnarray}
where $\mathbf{p}_{n}$, and $\mathbf{A}(t+\Delta t)$ denote the final
electron momenta and the vector potential at the time the electron leaves,
respectively. The integral extends over the ionization time $t^{\prime }$.
The $\delta $ function expresses the fact that the total kinetic energy of
the $N$ participating electrons is fixed by the first-ionized electron at
its recollision time $t$.

In the $3-N$ dimensional momentum space $(\mathbf{p}_{1},\ldots ,\ \mathbf{p}%
_{N}),$ the argument of the $\delta $ function corresponds to the equation
of a hypersphere
\begin{equation}
\frac{1}{2}\sum_{n=1}^{N}[p_{n\parallel }+A(t+\Delta t)]^{2}+\frac{1}{2}%
\sum_{n=1}^{N}\mathbf{p}{}_{n\perp }^{2}=E_{\mathrm{ret}}(t)-E_{0}^{(N)},
\label{hyper}
\end{equation}
where $p_{n\parallel }$ and $\mathbf{p}{}_{n\perp }(n=1,...,N)$ denote the
electron momentum components parallel and perpendicular to the laser-field
polarization, centered at $-A(t+\Delta t),$ and whose radius is determined
by the difference between the kinetic energy $E_{\mathrm{ret}}(t)$ of the
first electron upon return and the absolute value $E_{0}^{(N)}$ of the total
binding energy which must be overcome. Within the region delimited by this
hypersphere, in a classical framework, the process we are dealing with takes
place. Furthermore, if the transverse momenta are kept constant, Eq. (\ref
{hyper}) can be written as
\begin{equation}
\frac{1}{2}\sum_{n=1}^{N}[p_{n\parallel }+A(t+\Delta t)]^{2}=E_{\mathrm{ret}%
}(t)-\tilde{E}_{0}^{(N)},  \label{hypereff}
\end{equation}
where $\tilde{E}_{0}^{(N)}=E_{0}^{(N)}+\frac{1}{2}\sum_{n=1}^{N}\mathbf{p}%
{}_{n\perp }^{2}$ can be regarded as an effective binding energy
which the N electrons must overcome. Such an energy depends on the
transverse momentum components, and is minimal when they vanish$.$
By employing an adequate choice of field and atomic parameters, such
as, for instance, a few-cycle pulse, one can manipulate the radius
in (\ref{hyper}), making \ whole momentum regions collapse or
appear. In \cite{fewcycle1,fewcycle2}, we have exploited this in the
two-electron case in order to determine the absolute phase.

The only free parameter of this model is the time delay $\Delta t$ between
the recollision time and the time when multiple ionization occurs. It is the
sum of the time it takes to establish the statistical ensemble, i.e., the
thermalization time, and a possible additional ``dwell time'', until the
electrons become free. This model is an extension to NSMI of a classical
model introduced for NSDI in Refs.~\cite{FFetal04R,FFetal04} for $\Delta t=0$%
. Sufficiently high above threshold, it produced momentum distributions that
were virtually indistinguishable from their quantum-mechanical counterparts.

In order to integrate over unobserved momentum components, we exponentialize
the $\delta $ function in Eq.~(\ref{F}) using its Fourier representation
\begin{equation}
\delta (x)=\int_{-\infty }^{\infty }\frac{d\lambda }{2\pi }\exp (-i\lambda
x).  \label{delta}
\end{equation}
Infinite integrations over the momenta $\mathbf{p}_{n}$ can then be done by
Gaussian quadrature. The remaining integration over the variable $\lambda $
is performed employing \cite{GR}
\begin{equation}
\int_{-\infty }^{\infty }\frac{d\lambda }{(i\lambda +\epsilon )^{\nu }}%
e^{ip\lambda }=\frac{2\pi }{\Gamma (\nu )}p_{+}^{\nu -1},
\end{equation}
where $x_{+}^{\nu }=x^{\nu }\theta (x)$, with $\theta (x)$ the unit step
function and $\epsilon \rightarrow +0$.

The momentum distribution of the ion is then
\begin{eqnarray}
F(\mathbf{P}) &\equiv &\int \prod_{n=1}^{N}d^{3}\mathbf{p}_{n}\delta \left(
\mathbf{P}+\sum_{n=1}^{N}\mathbf{p}_{n}\right) F(\mathbf{p}_{1},\mathbf{p}%
_{2},\dots ,\mathbf{p}_{N})  \notag \\
&=&\frac{(2\pi )^{\frac{3N}{2}-\frac{3}{2}}}{N^{3/2}\Gamma (3(N-1)/2)}\int
dt^{\prime }R(t^{\prime })\left( \Delta E_{N,\mathrm{ion}}\right) _{+}^{%
\frac{3N}{2}-\frac{5}{2}}  \label{ion}
\end{eqnarray}
with $\Delta E_{N,\mathrm{ion}}\equiv E_{\mathrm{ret}}(t)-E_{0}^{(N)}-\frac{1%
}{2N}[\mathbf{P}-N\mathbf{A}(t+\Delta t)]^{2}$, where $\mathbf{P}$ gives the
momentum of the ion. In (\ref{ion}), we have used the momentum conservation $%
\mathbf{P}=-\sum_{n=1}^{N}\mathbf{p}_{n}$, which is valid in case the photon
momenta can be neglected.

If the ion momentum component perpendicular to the laser polarization is
entirely integrated over, the remaining distribution of the longitudinal ion
momentum $P_{\parallel }$ reads
\begin{eqnarray}
F(P_{\parallel }) &\equiv &\int d^{2}\mathbf{P}_{\perp }F(\mathbf{P})  \notag
\\
&=&\frac{(2\pi )^{\frac{3N}{2}-\frac{1}{2}}}{\sqrt{N}\Gamma ((3N-1)/2)}\int
dt^{\prime }R(t^{\prime })\left( \Delta E_{N,\mathrm{ion}\parallel }\right)
_{+}^{\frac{3N}{2}-\frac{3}{2}},  \label{ionpar}
\end{eqnarray}
where $\Delta E_{N,\mathrm{ion}\parallel }\equiv E_{\mathrm{ret}%
}(t)-E_{0}^{(N)}-\frac{1}{2N}[P_{\parallel }-NA(t+\Delta t)]^{2}$.

If just one transverse-momentum component (for instance, $P_{\perp ,2}$) is
integrated while the other one ($P_{\perp ,1}\equiv P_{\perp }$) is
observed, the corresponding distribution is
\begin{eqnarray}
F(P_{\parallel },P_{\perp }) &\equiv &\int dP_{\perp ,2}F(\mathbf{P})  \notag
\\
&=&\frac{(2\pi )^{\frac{3N}{2}-1}}{N\Gamma (3N/2-1)}\int dt^{\prime
}R(t^{\prime })\left( \Delta E_{N,\mathrm{ion}\parallel \perp }\right) _{+}^{%
\frac{3N}{2}-2}  \label{ionparperp}
\end{eqnarray}
with $\Delta E_{N,\mathrm{ion}\parallel \perp }\equiv \Delta E_{\mathrm{ion}%
\parallel }-\frac{1}{2N}P_{\perp }^{2}$. In the following section we will
explicitly compute ion momentum distributions employing our thermalization
model, i.e., from Eqs. (\ref{ionpar}) and (\ref{ionparperp}).

\section{Momentum distributions}

\label{results}

\subsection{Monochromatic fields}

As a starting point, we will consider a monochromatic laser field $\mathbf{E}%
(t)=-d\mathbf{A}(t)/dt$, for which the vector potential is of the form

\begin{equation}
\mathbf{A}(t)=A_{0}\cos (\omega t)\hat{e}_{z},
\end{equation}
with frequency $\omega $, amplitude $A_{0}$ and polarization axis $\hat{e}%
_{z}.$ This is a good approximation for long pulses. In this case,
$\Delta E_{N,\mathrm{ion}\parallel }(P_{||},t)=\Delta
E_{N,\mathrm{ion}\parallel }(-P_{||},t+T/2)$, where $T=2\pi /\omega
$ denotes a period of the driving
field, so that the momentum distributions are symmetric with respect to $%
P_{||}=0.$ Clearly, this also holds for $\Delta E_{N,\mathrm{ion}\parallel
\perp }.$ In our computations, we consider the shortest pair of trajectories
along which the first electron returns to its parent ion, and, in
particular, the individual contributions from the longer and the shorter
orbit within such a pair. Such orbits coalesce at the boundary of the region
determined by the hypershpere (\ref{hyper}) for which, classically,
nonsequential multiple ionization is allowed to occur. In the particular
model employed in this paper, such a boundary is determined by the minimal
momenta the $N$ electrons need in order to overcome the binding energy $%
E_{0}^{(N)}$, and the maximal momenta for which the first electron is able
to return.
\begin{figure}[tbp]
\includegraphics[width=7cm]{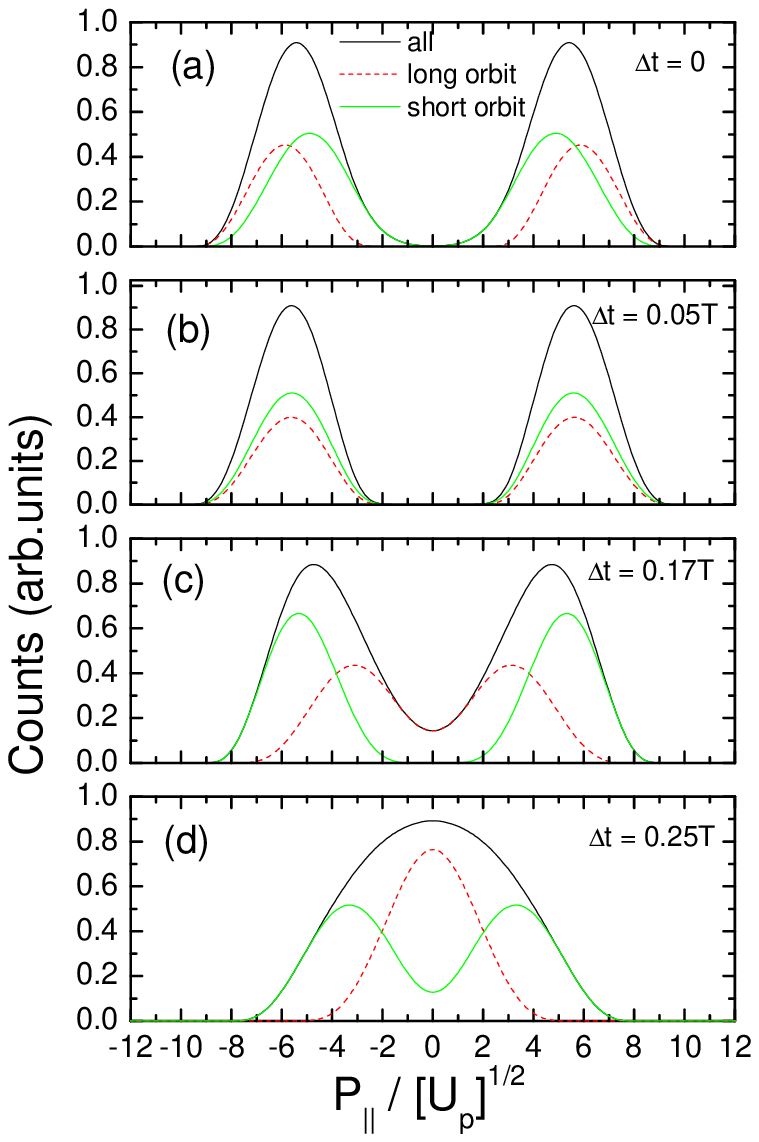}
\caption{(Color online) Distribution of the longitudinal ion momentum for
triple nonsequential ionization of neon at $I=2\times 10^{15}\mathrm{W/cm^{2}%
}$, for time delays $\Delta t=0$, $\Delta t=0.05T$, $\Delta t=0.17T$, and $%
\Delta t=0.25T,$ together with individual contributions from the
long and the short orbits. The ion momenta are given in units of
$[Up]^{1/2}$} \label{ne3int}
\end{figure}

In Fig.~\ref{ne3int}, we display the yield computed assuming that
the first electron returns to its parent ion either along the long
or the short orbit, together with the total momentum distributions,
for triple ionization of neon. In the upper panel
[Fig.~\ref{ne3int}.(a)], the thermalization time is taken to be
vanishing. As an overall feature, the total yield is approximately
peaked at $P_{||}=\pm 6\sqrt{U_{p}}$, as expected from classical
arguments. Furthermore, the shorter orbit yields distributions which
are slightly shifted towards lower momenta, in comparison to those
from the longer orbit. Such a shift is corrected if a time delay
$\Delta t=0.05T$ is introduced [Fig.~\ref{ne3int}.(b)]. This can be
easily understood in terms of the momentum transfer from the field
to the electrons. Such a momentum transfer is maximal if all
electrons leave at a crossing of the field, i.e., for $t+\Delta
t=n\pi $. For vanishing time delay, the electrons released by a
speed up collision, i.e., by an electron returning along the short
orbit, leave at roughly $t_{S}=0.95T$, while for
the slow-down collisions, this happens at $t_{L}=1.05T$. The time delay $%
\Delta t=0.05T$ moves the release times $t_{s}+\Delta t$ for the short-orbit
contributions towards the crossing and their long-orbit counterparts $%
t_{s}+\Delta t$ away from it. As a direct consequence, the contributions
from the former set of orbits move towards higher absolute momenta, while
those from the latter set move towards lower absolute momenta.

If the time delay increases [Fig.~\ref{ne3int}.(c)], the peak
momenta from the short-orbit contributions starts to move again
towards lower absolute
values. This is due to the fact that, for $\Delta t>0.05T$, the time $%
t_{S}+\Delta t$ all electrons leave starts to move away from the
crossing of the driving field. Hence, the drift momenta of the
electrons freed by the speed-up collisions decrease. One should
note, however, that the times $t_{L}+\Delta t$, for the electrons
released by the slow-down collisions, also move farther away from
the crossing. In this latter case, however, the peak momenta
decrease in a far more radical way. The fact that the two individual
contributions are now peaked at momenta which are very far apart
leads to a broadening in the momentum distributions. In this
context, it is worth mentioning that the time delay $\Delta t=0.17T$
yields the best agreement with the experimental findings, for
nonsequential triple and quadruple ionization \cite {multi1,multi2}.
This issue is discussed in more detail in \cite{prltherm}.

For exceptionally large time delays, such as those depicted in Fig. ~\ref
{ne3int}.(d), the contributions from the short orbit move back to lower
momenta, and those from the long orbit overlap at vanishing ion momenta. As
a direct consequence, the total yield merges in a single peak at vanishing
momentum. Such features resemble to a very large extent those obtained if
one considers sequential physical mechanisms, or the situation where the
electrons undergo a transition to an excited state, and, subsequently,
tunnel into the continuum. One should note, however, that for such a
parameter range, there is poor agreement with the experimental results.

\begin{figure}[tbp]
\includegraphics[width=8cm]{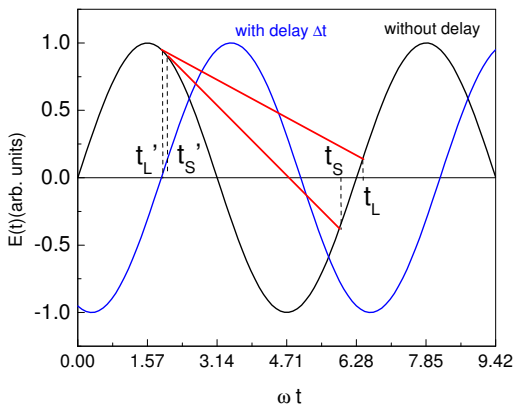}
\caption{(Color online) Start and return times for the long and
short orbits, denoted by $(t_{L}^{\prime },t_{L})$, $(t_{S}^{\prime
},t_{S})$, respectively, for vanishing and non-vanishing time delays
$\Delta t$. } \label{tanls}
\end{figure}

The effect observed in Fig. \ref{ne3int} is shown in more detail in Fig.~\ref
{tanls}, using the tangent construction mentioned in the previous section.
In the figure, we consider the return condition (\ref{tangent}) for the
short and long orbits. For vanishing time delays, the electron starting at $%
t_{S}^{\prime }$ and $t_{L}^{\prime }$ releases the remaining electrons at
the times $t_{S}$ before or $t_{L}$ after the crossing of the field,
respectively. Such times are determined by the intersection of the tangents (%
\ref{tangent}) with the dashed line. If, however, there is a time delay $%
\Delta t$, the time $t+\Delta t$ in which all electrons leave will be
determined by the intersection of the tangents (\ref{tangent}) with the
\textit{solid} line.
\begin{figure}[tbp]
\includegraphics[width=7cm]{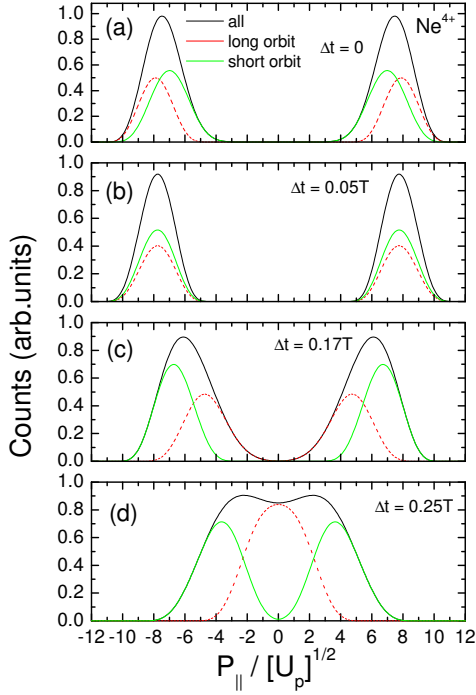}
\caption{(Color online) Distribution of the longitudinal ion momentum for
quadruple nonsequential ionization of neon at $I=2\times 10^{15}\mathrm{%
W/cm^{2}}$, for time delays $\Delta t=0$, $\Delta t=0.05T$, $\Delta t=0.17T$%
, and $\Delta t=0.25T,$ together with individual contributions from
the long and the short orbits. The ion momenta are given in units of
$[U_p]^{1/2}$.} \label{ne4int}
\end{figure}

In Fig. \ref{ne4int}, we depict the contributions from the slow-down and the
speed-up collisions, as well as from the total yield, for the
quadruple-ionization case and the same time delays as in Fig.~\ref{ne3int}.
 They exhibit the same overall behavior as in Fig.~\ref{ne3int}: the peak
momenta for the short-orbit contributions moves towards higher absolute
values for $\Delta t=0.05T$ \ [Fig. \ref{ne4int}.(b)], and, as $\Delta t$ is
further increased, the peak momenta start to decrease [Fig. \ref{ne4int}%
.(c)]. The maxima of the long-orbit contributions always shift towards $%
P_{\parallel }=0$, for increasing $\Delta t$, as shown in Fig. \ref{ne4int}%
.(b)-(d). This behavior is expected, since within our model, both triple and
quadruple nonsequential ionization are governed by the same physical
mechanism, namely energy transfer to all active electrons due to the
recollision of the first electron with its parent ion. The main difference
is the fact that all electrons must overcome a larger energy $E_{0}^{(4)}$
in order to reach the continuum. Therefore, for the same laser-field
intensity, as the charge state increases, the radius of the hypersphere (\ref
{hyper}) decreases and so does the momentum region for which the process in
question occurs. Furthermore, the peak momenta should increase to the
vicinity of $P_{||}=\pm 8\sqrt{U_{p}}.$ This leads, in general, to
distributions which are much more localized around the maxima than in the
triple-ionization case.

For a time delay $\Delta t=0.25T,$ however, instead of a single peak
at vanishing momenta, the total distribution exhibits a plateau,
with humps near its low-and high-energy ends [Fig. \ref{ne4int}
.(d)] . This is due to the fact that, while the region near
$P_{||}=0$ is filled by the contributions from the long orbit, the
contributions from the short orbit are still at a much larger
momenta than in the two electron case, and still cause such humps.
However, the distributions from the negative and positive momentum
regions still merge, as in the triple-ionization case.

\begin{figure}[tbp]
\includegraphics[width=6.5cm]{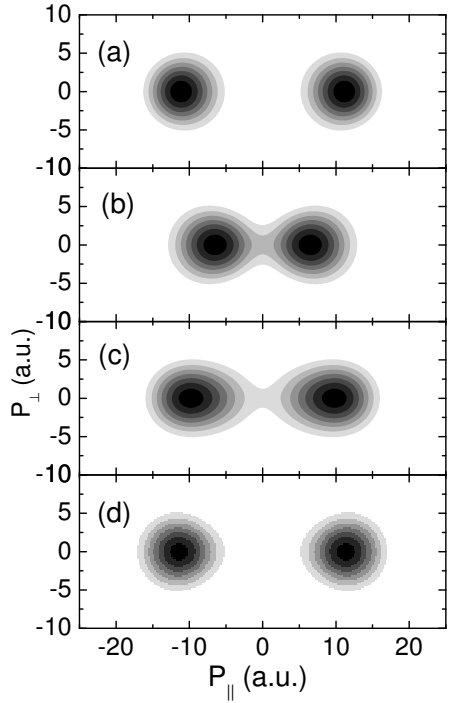}
\caption{Ion momentum distribution of nonsequential triple
ionization of neon at $I=1.5\times 10^{15}\mathrm{W/cm^{2}}$, and
time delay $\Delta t=0.17T$. Panels (a), (b) and (c) correspond to
the contributions of the short orbit, of the long orbit, and to the
total yield, respectively. In panel (d), the total yield for
vanishing time delay is given for comparison. The ion momenta are
given in atomic units.} \label{ne3diff}
\end{figure}
Finally, in Fig. \ref{ne3diff}, we analyze the individual contributions of
both orbits if the transverse momenta are only partly integrated over, for
time delays $\Delta t=0$ and $\Delta t=0.17T$. \ The distributions from the
short and the long orbits exhibit a displacement in their peak momenta which
is similar to those obtained in the non-resolved case. In particular, the
elongation observed in experiments for such distributions is directly
related to this displacement.

\subsection{Few-cycle pulses}

In this section, we will take the driving field to be a few-cycle pulse $%
\mathbf{E}(t)=-d\mathbf{A}(t)/dt$, corresponding to the vector potential

\begin{equation}
\mathbf{A}(t)=A_{0}\sin ^{2}[\omega t/(2n)]\sin (\omega t+\phi )\hat{e}_{z}
\label{pulse}
\end{equation}
where $\phi $ is the absolute phase, and $n$ is the number of cycles. We
choose a four-cycle pulse $(n=4)$, and consider triple and quadruple
ionization. Apart from their wide range of applications (for reviews see,
e.g., \cite{fewcyclerevs}), few-cycle pulses are very attractive in the
context of nonsequential multiple ionization. This is due to the fact that
several physical mechanisms which would compete with electron-impact
ionization, such as excitation-tunneling ionization, are strongly suppressed
in the few-cycle case \cite{multi1}. Therefore, in principle, one expects
cleaner measurements.
\begin{figure}[tbp]
\includegraphics[width=11cm]{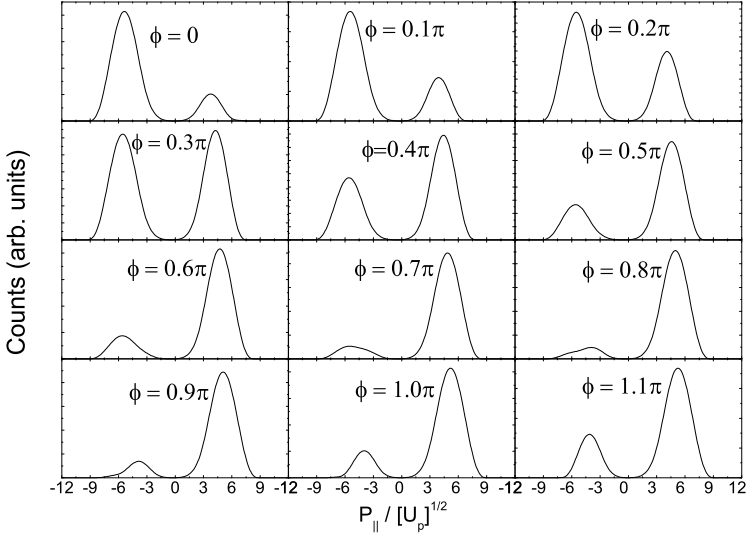}
\caption{Ion momentum distribution of nonsequential triple ionization of
neon by a four cycle pulse of intensity $I=2\times 10^{15}\mathrm{W/cm^{2}}$%
, and several absolute phases $\protect\phi .$ We assume that the
electrons leave instantaneously after recollision ($\Delta t=0)$.
The ion momenta are given in units of $[U_p]^{1/2}$.}
\label{ne3_CEphase}
\end{figure}

In Fig. \ref{ne3_CEphase}, we present ion momentum distributions for
triple ionization of neon, computed with the driving field
(\ref{pulse}). As a starting point, we take the thermalization time
$\Delta t$ to be vanishing. In contrast to the monochromatic case,
such distributions are in general asymmetric with respect to
$P_{\parallel }=0.$ This is expected since, for a few cycle pulse,
$A(t)=\pm A(t+nT/2)$ \ holds, if at all, only for very specific sets
of parameters. Therefore
the relation $\Delta E_{N,\mathrm{ion}\parallel }(P_{||},t)=\Delta E_{N,%
\mathrm{ion}\parallel }(-P_{||},t+T/2)$ is in general not fulfilled.
Depending on the absolute phase, the distributions are strongly
localized in either the positive or negative momentum region. Around
a critical phase, they start to shift from one momentum region to
the other.

Such a behavior is very similar to that observed for nonsequential
double ionization ($N=2$). In the two-electron case, this shift has
been employed to diagnose the absolute phase \cite
{fewcycle1,fewcycle2,fewcycleexp}. Furthermore, it has been
explained as a change in the dominant set of trajectories for an
electron rescattering with
its parent ion. Such a set is determined by the interplay between the rate (%
\ref{quasistatic}) with which the first electron is released in the
continuum, and the available phase space: if there was a large
probability for the first electron to leave along a particular set
of orbits, and if it returned with enough energy to cause
appreciable electron-impact ionization, one expects the
contributions from such orbits to the yield to be large.

\begin{figure}[tbp]
\includegraphics[width=12cm]{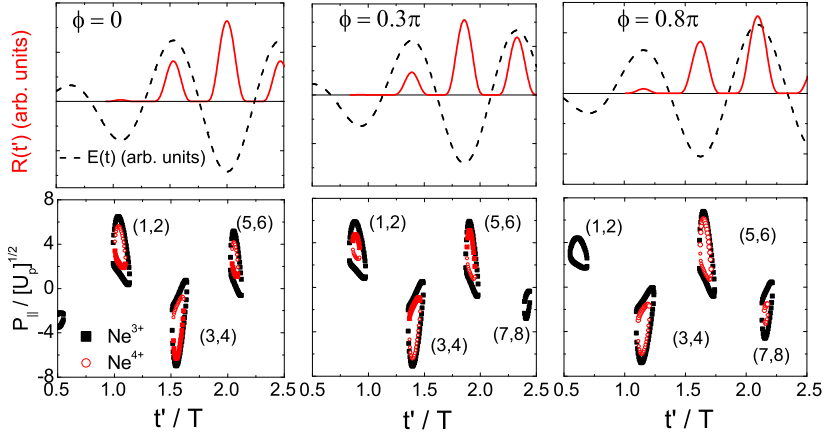}
\caption{(Color online) Quasi-static tunneling rate (upper panels), and
parallel momenta $P_{\parallel }$ (lower panels), as functions of the start
times $t^{\prime }$, for triple and quadruple ionization of neon. We
consider absolute phases $\protect\phi =0$, $\protect\phi =0.3\protect\pi $
and $\protect\phi =0.8\protect\pi ,$ the same driving-field parameters as in
the previous figure, vanishing transverse momenta $P_{\perp }$ and
thermalization times $\Delta t=0$. The numbers $(i,j)$ in the figure
indicate a pair of orbits, with the odd and even numbers referring to the
longer and the shorter orbit in a pair, respectively. The start time $%
t^{\prime }$ is given in units of the field cycle and the ion
momenta in units of $[U_{p}]^{1/2}.$} \label{phasesp}
\end{figure}

Within the present model, this explanation also holds for $N>2,$ and
is illustrated in Fig. \ref{phasesp} for the parameter set of Fig.
\ref {ne3_CEphase}. In the upper and lower panels, we present the
quasi-static tunneling rate and parallel momentum components for the
residual ion, respectively, as functions of the emission time
$t^{\prime }.$ In the latter case, we consider the transverse ionic
momentum components to be vanishing. This gives an upper bound for
the momentum region for which electron-impact ionization is allowed
to occur, since, in Eq.~(\ref{hypereff}), the effective binding
energy $\tilde{E}_{0}^{(N)}$ is minimal.

In the figure, we identify at least three sets of orbits, whose start times $%
t^{\prime }$ are highly dependent on the carrier-envelope phase
$\phi .$ For instance, for $\phi =0$, there exist three such pairs:
$(1,2)$, $(3,4)$ and $(5,6)$. Thereby, the odd and even numbers
refer to the longer and the shorter orbit in a pair, respectively.
Orbits $(1,2)$ and $(5,6)$ contribute to the yield in the positive
momentum region, whereas $(3,4)$ leads to a peak in the region
$P_{\parallel }<0.$ Since, for $(1,2)$, the quasi-static rate (\ref
{quasistatic}) is very small, the contributions from this set of
orbits is negligible. Hence, the features observed in Fig. 5 will be
determined by the remaining two pairs. In fact, orbits $(3,4)$ and
$(5,6)$ lead to the large peak for negative momenta and to the small
peak for positive momenta, respectively. For the set $(3,4)$, there
is a large momentum region for which the first electron may dislodge
the remaining two, and a non-negligible ionization rate. Therefore,
its contributions to the yield are expected to be quite prominent.
For the set $(5,6)$, even though the tunneling rate is very large,
there is only a small region for which electron-impact ionization is
allowed. Hence, the pair $(3,4)$ dominates.

 This picture starts to
change around the critical phase $\phi _{c}=0.3\pi ,$ for which the
yield is approximately symmetric. In comparison to the case $\phi
=0$, there is now a decrease in the quasi-static rate for $(3,4)$
and an increase in the momentum region for $(5,6)$. Therefore, the
contributions from the former orbits decrease and those from the
latter orbits increase, respectively. Beyond the critical phase (for
instance, for $\phi=0.8\pi$), the yield is mainly concentrated in
the positive momentum region, since now $(5,6)$ prevail. For this
latter case, there is also a further set of orbits which contribute
to the yield in the region $P_{\parallel }<0$, namely $(7,8)$. For
such a set, the rate (\ref {quasistatic}) is very large. However,
the region for which electron-impact ionization is allowed to occur
is very small and $(5,6)$ is still the dominant pair of orbits.

For a higher charge state, the very same line of argumentation
applies. The main difference is that the first electron will need to
overcome a larger binding energy in order to release the remaining
electrons. As a direct consequence, the radius of the hypersphere
(\ref{hyper}) will be smaller so that whole momentum regions may
collapse, or contribute in a much less pronounced way to the yield.
For instance, for nonsequential quadruple ionization, for the
absolute phase $\phi =0$ the small peak on the positive momentum
region is absent. This is due to the fact that the momentum space
for the orbits $(5,6)$ is much smaller than in the three-electron
case, so that their contributions to the yield are negligible. For
$\phi=0.8\pi$, we have also observed that the small peak in the
negative momentum region is not present, for the same reason.

\begin{figure}[tbp]
\includegraphics[width=10cm]{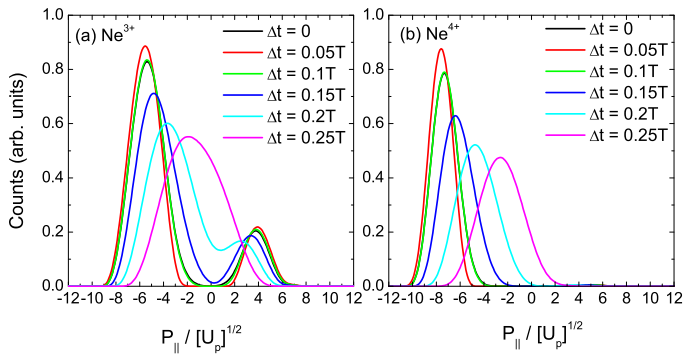}
\caption{(Color online) Distribution of the longitudinal ion momentum for
triple [panel (a)] and quadruple [panel (b)] nonsequential ionization of neon%
$,$for a four-cycle pulse of intensity at $I=2\times
10^{15}\mathrm{W/cm^{2}} $, absolute phase $\protect\phi =0,$ and
several time delays. The momenta are given in units of
$[U_p]^{1/2}$.} \label{delays}
\end{figure}

We now consider non-vanishing thermalization times and a constant
absolute phase, which has been chosen so that the yield is almost
entirely concentrated in the negative momentum region. The results
of these computations are displayed in Fig. \ref{delays}. As an
overall feature, the time delays $\Delta t$ influence the width and
the position of the peaks of the distributions. They do not modify,
however, their asymmetric shapes or the momentum region the yield is
concentrated. This is due to the fact that a non-vanishing $\Delta
t$ changes the drift momenta the $N$ electrons acquire from the
external laser field when they reach the continuum, i.e., the
left-hand side in Eqs. (\ref{hyper}) and (\ref{hypereff}), but does
not affect the quasi-static tunneling rate (\ref{quasistatic}) or
the binding energy the electrons must overcome [i.e., the right-hand
side in Eqs. (\ref {hyper}), (\ref{hypereff})]. Therefore the
dominant pairs of orbits remain the same. In other words, by
changing $\Delta t$ one influences the center of the hypersphere
(\ref{hyper}), but not its radius.

\begin{figure}[tbp]
\includegraphics[width=7.5cm]{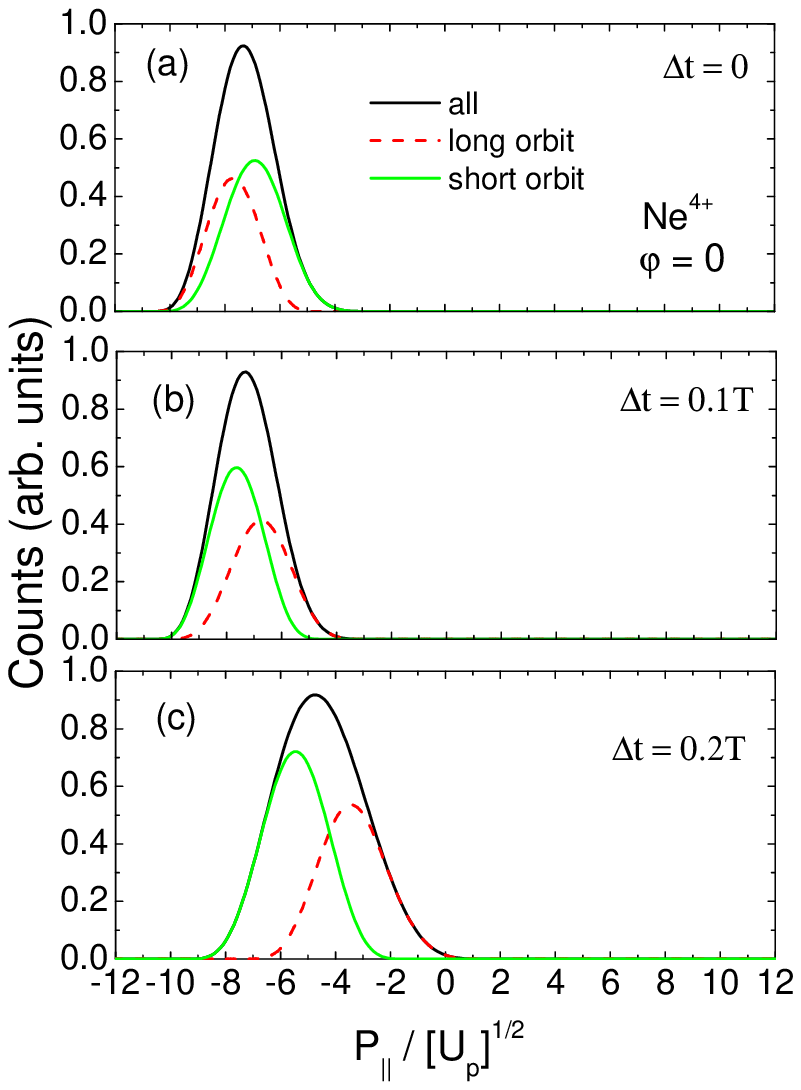}
\caption{(Color online) Distribution of the longitudinal ion momentum
quadruple nonsequential ionization of neon$,$together with the individual
contributions from the long and short orbits, for a four-cycle pulse of
intensity $I=2\times 10^{15}\mathrm{W/cm^{2}}$, absolute phase $\protect\phi %
=0.$ We consider the time delays $\Delta t=0,$ $\Delta t=0.1T$ and
$\Delta t=0.2T.$ The ion momenta are given in units of
$[U_p]^{1/2}$. } \label{fewcyc3}
\end{figure}

We shall now have a closer look at how the width of the
distributions is affected by the interplay between the long and the
\ short orbit. In Fig. \ref{fewcyc3}, we display the ion momentum
distributions for quadruple ionization in the few-cycle pulse case,
together with the individual contributions from the long and the
short orbits. For a few-cycle pulse, we define such orbits with
respect to the pair which gives dominant contributions to the yield.
For this specific parameter set, this means orbits $(3,4)$ in Fig.
\ref{phasesp}. The long- and short-orbit contributions, i.e., the
individual contributions from orbits $3$ and $4$, respectively,
behave in a similar way to the monochromatic-field case (Figs.
\ref{ne3int} and \ref{ne4int}). Indeed, for vanishing thermalization
times, the momentum distributions from the slow-down collisions are
peaked at a
slightly larger absolute momentum than those from the short orbit. For $%
\Delta t=0.1T$ [Fig. \ref{fewcyc3}.(b)], the former and the latter
contributions have moved away from or towards $P_{\parallel }=0$,
respectively. Once more, as the thermalization time increases, the
peaks of both contributions shift towards vanishing ion momenta, as
shown in Fig.~\ref{fewcyc3}.(c).

A qualitative difference, however, between Fig. \ref{fewcyc3} and
its monochromatic counterpart is that, for large time delays, one no
longer obtains a plateau in the vicinity of vanishing ion momenta,
but a single peak in the negative momentum region. This is due to
the fact that, for a monochromatic driving field, the symmetric
peaks in the positive and negative momentum regions start to merge.
For a few-cycle driving pulse, on the other hand, by taking an
appropriate absolute phase, the distributions can be chosen in such
a way that they are concentrated only in one momentum region. Hence,
it is possible to avoid that a plateau or a single peak occurs near
$P_{\parallel }=0.$

\section{Conclusions}

\label{conclusions}

We have investigated the influence of the shorter and the longer
orbit of an electron recolliding inelastically with its parent ion
on ion momentum distributions, within the context of a
thermalization model for laser-induced $N-$fold nonsequential
ionization. In this model, the first electron, upon recollision,
shares the kinetic energy it acquired from the external laser field
with the remaining $N-1$ bound electrons. All $N$ electrons then
leave after a time interval $\Delta t$. We have taken the driving
field to be a monochromatic wave and a few-cycle laser pulse.

We have traced the variations in the widths and peaks of such
momentum distributions, with respect to the time delays $\Delta t$,
back to the momentum transfer from the field to an electron
traveling along the short or the long orbit, and recolliding with
its parent ion. The momentum transfer is maximal at a crossing of
the electric field. With increasing time delays, the time for which
the electrons released by a slow-down collision, i.e., by an
electron moving along the long orbit, will leave their parent ion
moves away from this crossing. As a direct consequence, the momentum
transfer will decrease and the peak momenta will move towards lower
absolute values. The situation is however quite different for the
electrons released by a speed up collision. In this case, following
the increase of time delay, the release time $t+\Delta t$ will
initially (i.e., for $\Delta t\leq 0.05T$) approach a crossing, and
subsequently (i.e., for $\Delta t>0.05T$) move away from it. This
will make the peaks of the pertinent momentum distributions move to
larger absolute values, and then back towards $P_{\parallel }=0$.
The width of the resulting total momentum distributions, as well as
their peaks, will result from the interplay between such trends.
This also explains the elongated patterns observed for the momentum
distributions, in case the perpendicular momentum of the ion is
resolved.

In both monochromatic and few-cycle cases, we considered the dominant pair
of orbits, which, for each situation, is chosen in a slightly different way.
For monochromatic fields, we took the two shortest trajectories, i.e., those
for which the first electron spends the shortest time in the continuum.
Physically, this pair is dominant, since the spreading of the electronic
wave packet is less pronounced than for longer pairs of orbits. Hence, the
overlap with bound-state wave packets will be appreciable\footnote{%
We would like to stress out that wave-packet spreading is not
included in our classical model. It is, however, present in the
quantum-mechanical model we have employed in the two-electron case,
which is its quantum-mechanical counterpart. In this latter case and
as expected from our physical intuition, we verified that the main
contributions to the yield come from the shortest pair.}. In the
few-cycle case, the dominant pair is determined by the interplay
between the tunneling rate (\ref{quasistatic}) for the first
electron and the available momentum space. A pair of orbits will
only contribute in a relevant way to the yield if the first electron
is ejected with a large probability and if, upon return, it has
enough energy to release the second electron (see, e.g.
\cite{fewcycle1,fewcycle2}, for more details).

Specifically for few-cycle pulses, we have shown that non-vanishing
thermalization times do not play a very important role in
determining the shape of \ the momentum distributions. This is due
to the fact that a time delay $\Delta t\neq 0$ between recollision
and multiple ionization only alters the drift momenta the $N$\
electrons acquire from the field when they reach the continuum.
However, this delay does not modify the key ingredients which
determine the dominant pairs of orbits and therefore the shape of
the momentum distributions: the tunneling rate for the first
electron and its kinetic energy upon return. In particular, both
ingredients are highly dependent on the carrier-envelope phase, so
that the shape of the momentum distributions can be used for
measuring this parameter \cite {fewcycle1,fewcycleexp}. This means
that, even in case there is a non-vanishing thermalization time, one
could, in principle, still employ nonsequential multiple ionization
for absolute-phase diagnosis. Finally, few-cycle driving pulses may
be very useful for testing and validating the thermalization model
further, since many other ionization channels, such as
excitation-tunneling or sequential ionization, are strongly
suppressed in this case.

\vspace*{0.5cm} \noindent\textbf{Acknowledgements:} The authors
would like to thank W. Becker for useful discussions. C.F.M.F. is
also grateful to R. Moshammer and especially to A. Rudenko for
discussions on nonsequential multiple ionization with few-cycle
pulses, and to the Max Planck Institut f\"{u}r Kernphysik,
Heidelberg, for its kind hospitality.

\end{document}